\documentclass[runningheads]{llncs}
\usepackage[T1]{fontenc}
\usepackage{hyperref}
\usepackage{graphicx}
\begin{document}
\title{xLSTM-UNet can be an Effective 2D \& 3D 
Medical Image Segmentation Backbone with Vision-LSTM (ViL) better than its Mamba Counterpart}
\titlerunning{xLSTM-UNet in 2D \& 3D Biomedical Image Segmentation}
%
\author{Tianrun Chen\inst{1,2}$^{+}$,
Chaotao Ding\inst{2,3}$^{+}$, Lanyun Zhu\inst{4}$^{+}$, Tao Xu\inst{3}$^{+}$, Yan Wang\inst{7}, Deyi Ji\inst{6}, Ying Zang\inst{3}$^{*}$, Zejian Li\inst{5}}

\authorrunning{T, Chen et al.}

\institute{College of Computer Science and Technology, Zhejiang University.  \and \href{http://www.kokoni3d.com/}{KOKONI, Moxin (Huzhou) Tech. Co., LTD.} \and
School of Information Engineering, Huzhou University. \and 
Information Systems Technology
and Design Pillar, Singapore University of Technology and Design.\and
School of Software Technology, Zhejiang University. \and
School of Information Science and Technology, University of Science and Technology of China.\and
School of Instrumentation and Optoelectronic Engineering; State Key Laboratory of Software Development Environment, Beihang University.\\ 
+ Equal Contribution    * Corresponding Author \\
\email{tianrun.chen@zju.edu.cn; 02750@zjhu.edu.cn}\\
Project Page: \href{http://tianrun-chen.github.io/xLSTM-UNet/}{http://tianrun-chen.github.io/xLSTM-Unet/}\\
\textbf{TL; DR: We replace Mamba with xLSTM in UMamba. It works better!}
}

\maketitle              
\begin{abstract}
Convolutional Neural Networks (CNNs) and Vision Transformers (ViT) have been pivotal in biomedical image segmentation. Yet, their ability to manage long-range dependencies remains constrained by inherent locality and computational overhead. 
To overcome these challenges, in this technical report, we first propose xLSTM-UNet, a UNet structured deep learning neural network that leverages Vision-LSTM (xLSTM) as its backbone for medical image segmentation. xLSTM has recently been proposed as the successor of Long Short-Term Memory (LSTM) networks and has demonstrated superior performance compared to Transformers and State Space Models (SSMs) like Mamba in Neural Language Processing (NLP) and image classification (as demonstrated in Vision-LSTM, or ViL implementation). Here, xLSTM-UNet we designed to extend the success in the biomedical image segmentation domain. By integrating the local feature extraction strengths of convolutional layers with the long-range dependency-capturing abilities of xLSTM, xLSTM-UNet offers a robust solution for comprehensive image analysis. We validate the efficacy of xLSTM-UNet through experiments. Our findings demonstrate that xLSTM-UNet consistently surpasses the performance of leading CNN-based, Transformer-based, and Mamba-based segmentation networks in multiple datasets in biomedical segmentation including organs in abdomen MRI, instruments in endoscopic images, and cells in microscopic images. With comprehensive experiments performed, this technical report highlights the potential of xLSTM-based architectures in advancing biomedical image analysis in both 2D and 3D. The code, models, and datasets are publicly available at \href{http://tianrun-chen.github.io/xLSTM-UNet/}{http://tianrun-chen.github.io/xLSTM-Unet/}.

\end{abstract}
\section{Introduction}
Biomedical image segmentation is a critical task in medical imaging, enabling precise delineation of anatomical structures and anomalies essential for diagnosis, treatment planning, and research \cite{ma2023towards,ma2024segment}. In recent years, deep learning methods have achieved remarkable success in tumor segmentation \cite{bilic2023liver,heller2021state} and organ segmentation in 3D Computed Tomography (CT) scans \cite{ma2023unleashing}, as well as in cell segmentation in microscopy images \cite{stringer2021cellpose,meijering2012cell,ma2024multimodality,gupta2023segpc}. These advancements underscore the transformative impact of deep learning on the landscape of biomedical image segmentation, paving the way for more accurate and efficient diagnostic and treatment planning tools. Traditionally, Convolutional Neural Networks (CNNs) have been the backbone of this domain in deep learning-enabled methods, leveraging their powerful local feature extraction capabilities \cite{ronneberger2015u,milletari2016v,zhou2018unet++,oktay2018attention,huang2020unet}. More recently, Vision Transformers (ViTs) have gained popularity by offering a robust alternative, capable of capturing global context through self-attention mechanisms \cite{chen2021transunet,wang2023swinmm,hatamizadeh2021swin,hatamizadeh2022unetr,lin2022ds,shaker2024unetr++,shamshad2023transformers,azad2023advances,you2022class,yu2023unest}. Despite their successes, both CNNs and ViTs face inherent limitations. CNNs struggle with long-range dependencies due to their localized receptive fields, while ViTs encounter substantial computational overhead \cite{pang2023slim,yuan2023effective,he2023h2former,xie2021cotr}, especially with high-resolution images or high-dimensional imaging modalities like 3D images or hyper-spectral imaging like stimulated Raman scattering (SRS) imaging \cite{zhang2011highly,zhang2014fast,zhang2015coherent,chen2023high,chen2021machine} or mid-infrared (IR) spectroscopic imaging \cite{fu2020pushing,fu2023super,zhang2016depth}.

To address these challenges, recent work has proposed to integrate computation modules that have long-range dependencies and also exhibit linear computational and memory complexity w.r.t. sequence length. Among these computation modules, State Space Models (SSMs) \cite{kalman1960new,gu2021combining,gu2023modeling}, like Mamba \cite{gu2023mamba}, has demonstrated its huge success. SSMs excel in handling long-range dependencies and have been successfully integrated into conventional UNet architectures. Variants like UMamba \cite{ma2024u}, VM-Unet\cite{ruan2024vm,zhang2024vm,wu2024ultralight}, Mamba-Unet \cite{wang2024mamba}, Swin-UMamba \cite{liu2024swin}, and SegMamba \cite{xing2024segmamba}, have demonstrated their considerable success.

Meanwhile, Extended Long-Short Term Memory (xLSTM) has recently emerged as a powerful successor to Long Short-Term Memory (LSTM) networks, challenging Transformers in sequence modeling \cite{beck2024xlstm}. Like SSMs, xLSTM can handle long-range dependencies and maintain linear computational and memory complexity. However, xLSTM has demonstrated superior performance in neural language processing (NLP) and image classification (in its Vision-LSTM (ViL) implementation \cite{alkin2024vision}). This success naturally raises the question: \textbf{Can xLSTM, or ViL, also excel in image segmentation, specifically in the field of medical image segmentation?}

The answer is \textbf{\textit{Yes!}} In this technical report, we introduce xLSTM-UNet, the first xLSTM-enabled U-Net image segmentation network that can perform both 2D and 3D medical image segmentation tasks and achieves state-of-the-art (SOTA) results. We conducted comprehensive experiments in various 2D and 3D medical segmentation scenarios, including organs in abdominal MRI, instruments in endoscopy, cells in microscopy, and cancer segmentation in 3D brain MRI volumes. The results show that xLSTM-UNet outperforms existing CNN-based and Transformer-based segmentation methods, as well as its Mamba-based counterparts. These findings highlight the potential of xLSTM-based architecture to set new benchmarks in the field of medical image segmentation, offering improved accuracy and efficiency across a wide range of applications. To further advance research in this area, we will release the model and code in \href{http://tianrun-chen.github.io/xLSTM-UNet/}{http://tianrun-chen.github.io/xLSTM-Unet/}, enabling future explorations in various fields such as automated pathology detection, camouflaged image segmentation, precision agriculture, environmental monitoring, satellite imagery analysis, and industrial inspection, and so on. 

\section{Method}
\textbf{TL; DR: We just replaced Mamba with xLSTM blocks in Mamba-based UNet network \cite{ma2024u,xing2024segmamba} and it works.}
\begin{figure}[t]
\includegraphics[width=\textwidth]{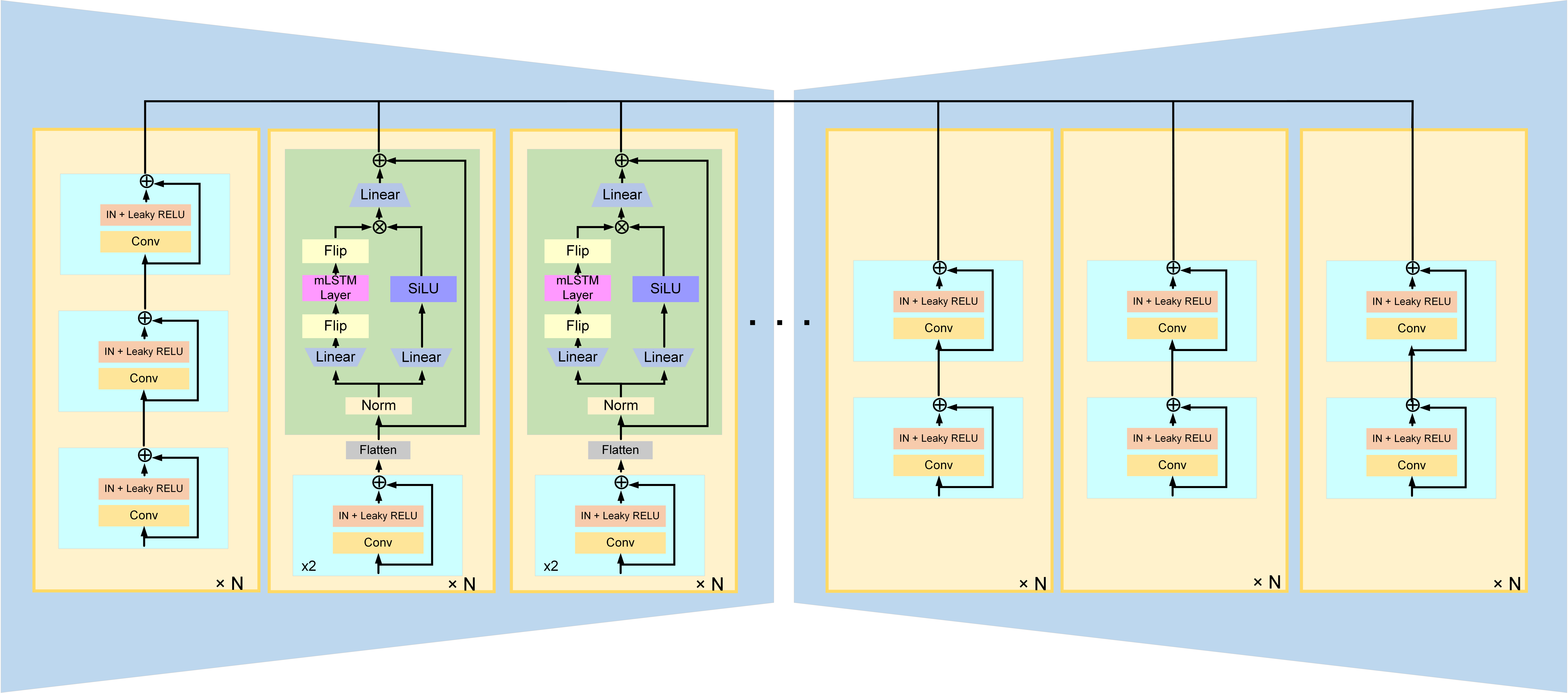}
\caption{The framework of the proposed method.} \label{fig1}
\end{figure}

Fig. \ref{fig1} showcases the xLSTM-UNet network architecture. xLSTM-UNet follows a conventional UNet-like structure. The input information first passes a convolution layer for initial down-sampling. Then several subsequent layers that are constructed using the aforementioned xLSTM building blocks to capture both local features and long-range dependencies form the main part of the encoder. Note that the xLSTM-UNet is designed with the goal of harnessing the best aspects of both U-Net and xLSTM for improved global comprehension in medical image understanding. Therefore, instead of only applying xLSTM in the compressed latent space after the down-sampling has finished, we hereby use the xLSTM in multiple layers in the encoder, in which each layer contains two successive Residual blocks with one plain convolution and an Instance Normalization (IN) and followed by a xLSTM block as in \cite{alkin2024vision}. Specifically, the image feature passing the residual blocks has a shape of $(B, C, H, W, D)$, which is first flattened and transposed to $(B, H \times W \times D, C)$, followed by a layer normalization, and then feed to the ViL block. Such the practice of involving xLSTM in multiple layers helps the feature extractions in multiple resolutions/perception fields, and this information extracted by xLSTM blocks is reshaped to $(B, C, H, W, D)$ and concatenated to the layers in the decoding steps to facilitate the segmentation mask generation. 

After encoding, the decoder, comprising Residual blocks and transposed convolutions, concentrates on the meticulous recovery of detailed local information. Additionally, we inherit the skip connection from the U-Net architecture to interconnect the hierarchical features from the encoder to the decoder. The final decoder feature is fed into a 1 x 1 convolutional layer, coupled with a Softmax layer, to generate the ultimate segmentation probability map. Furthermore, following \cite{ma2024u}, we also implemented a variant where the U-xLSTM block is exclusively utilized in the bottleneck, denoted as 'ours\_bot', while 'ours\_enc,' denotes the network that applies xLSTM block in all encoder blocks.

\section{Experiments}
\subsection{Datasets}
\textbf{TL; DR: We use the same benchmarks as UMamba \cite{ma2024u} for 2D medical image segmentation, and the same benchmarks as SegMamba for 3D medical image segmentation.\cite{xing2024segmamba}.}

To validate the effectiveness of our proposed method (xLSTM-UNet), we utilized several representative medical image segmentation datasets. These datasets encompass organ segmentation, instrument segmentation, and cell segmentation, and they span various resolutions and image modalities. By employing these diverse datasets, we can comprehensively assess the performance and applicability of xLSTM-UNet in different scenarios, thereby demonstrating its efficacy and superiority in medical image segmentation.

 \textbf{Abdomen MRI:} We utilized the Abdomen MRI dataset from the MICCAI 2022 AMOS Challenge \cite{ma2023unleashing}, which is dedicated to the segmentation of abdominal organs. The dataset was meticulously annotated by radiologists using MedSAM \cite{ma2024segment} and ITK-SNAP \cite{yushkevich2016itk}. In accordance with the U-Mamba \cite{ma2024u} settings, we employed 60 labeled MRI scans for training and 50 MRI scans for testing, comprising 5615 slices in the training set and 3357 slices in the testing set. This dataset encompasses 13 distinct class labels: liver, spleen, pancreas, right kidney, left kidney, stomach, gallbladder, esophagus, aorta, inferior vena cava, right adrenal gland, left adrenal gland, and duodenum. For training purposes, in 2D segmentation tasks, the images were cropped into patches with a resolution of 320 x 320 pixels. For 3D segmentation tasks, the patches were set to a size of 48 x 160 x 224 pixels.

\textbf{Endoscopy images:} The endoscopy image dataset was sourced from the MICCAI 2017 EndoVis Challenge\cite{allan20192017}, focusing on the segmentation of seven surgical instruments from endoscopic images. These instruments include the large needle driver, prograsp forceps, monopolar curved scissors, cadiere forceps, bipolar forceps, vessel sealer, and an additional drop-in ultrasound probe. We adhered to the official dataset split, which comprises 1800 image frames for the training set and 1200 image frames for the testing set. The training images were extracted from eight videos, while the testing set consisted of unseen images from two new videos. The images were cropped to a size of (384, 640) pixels to accommodate the nnU-Net framework for both training and testing.

\textbf{Microscopy images:} The microscopy image dataset was obtained from the NeurIPS 2022 Cell Segmentation Challenge\cite{ma2024multimodality}, which focuses on cell segmentation in various microscopy images. We used 1000 images for training and 101 images for evaluation. The original task was an instance segmentation task. In our experiments, we converted instance segmentation into a semantic segmentation task following the data processing method described in \cite{ma2024u}. The images were cropped to a size of (512, 512) pixels to accommodate the nnU-Net framework for both training and testing.

\textbf{BraTS2023:} The BraTS2023 dataset\cite{menze2014multimodal,kazerooni2023brain,bakas2017advancing} comprises 1,251 3D brain MRI volumes. Each volume features four imaging modalities (T1, T1Gd, T2, and T2-FLAIR) and three segmentation targets: Whole Tumor (WT), Enhancing Tumor (ET), and Tumor Core (TC). For training, we use a random crop size of 128×128×128 to process the 3D data.

\subsection{Implemetation details}
The network is implemented based on UMamba \cite{ma2024u}. The loss function used is the sum of Dice loss and cross-entropy loss. We employ the AdamW optimizer with a weight decay of 0.05. The learning rates were set to 0.005 for the Abdomen MRI dataset, 0.01 for the Endoscopy dataset, 0.007 for training xLSTM-UNet\_Bot on the Microscopy dataset, 0.0015 for training xLSTM-UNet\_Enc on the Microscopy dataset, and 0.01 for training on the BraTS2023 dataset. The batch sizes were set as follows: 2 for the 3D Abdomen MRI dataset, 30 for the 2D Abdomen MRI dataset, 2 for the Endoscopy dataset, 12 for the Microscopy dataset, and 4 for the BraTS2023 dataset. All networks were trained from scratch for 1000 epochs on a single NVIDIA A100 GPU. For more implementation details, please refer to our codebase.

\subsection{Baselines}
\textbf{TL; DR: The baseline is the same as that in the UMamba and SegMamba models \cite{ma2024u,xing2024segmamba} (and we outperform them!)}

In 2D medical segmentation, to ensure a fair comparison, we follow the evaluation protocol in UMamba \cite{ma2024u}. We selected two CNN-based segmentation networks (nnU-Net \cite{ma2023unleashing} and SegResNet \cite{myronenko20193d}) and two Transformer-based networks (UNETR \cite{hatamizadeh2022unetr} and SwinUNETR \cite{hatamizadeh2021swin}), as well as the UMamba itself, which has two variations: U-Mamba\_Bot and U-Mamba\_Enc. Similar to our configuration, U-Mamba\_Bot is applied only at the bottleneck, while U-Mamba\_Enc is used in each encoder. We used the Dice Similarity Coefficient (DSC) and Normalized Surface Distance (NSD) as evaluation metrics for the semantic segmentation tasks on the Abdomen MRI and Endoscopy datasets \cite{maier2024metrics}. For cell segmentation on the Microscopy dataset, we employed the F1 score.

In 3D medical segmentation, for the 3D Abdomen MRI dataset, the baseline methods and tasks remain consistent with those used in 2D medical segmentation. For the BraTS2023 dataset, to ensure a fair comparison, we follow the evaluation protocol outlined in SegMamba \cite{xing2024segmamba}. We use the same baseline methods, including three CNN-based methods (SegresNet \cite{myronenko20193d}, UX-Net \cite{lee20223d}, MedNeXt \cite{roy2023mednext}), three transformer-based methods (UNETR \cite{hatamizadeh2022unetr}, SwinUNETR \cite{hatamizadeh2021swin}, and SwinUNETR V2 \cite{he2023swinunetr}), and the Mamba-based method SegMamba itself. Dice and HD95 were used as evaluation metrics.

\subsection{Quantitative and Qualitative Results for 2D Segmentation}

Table \ref{table:1} presents the segmentation performance of various methods on the Abdomen MRI 2D, Endoscopy, and Microscopy datasets. Our proposed xLSTM-UNet outperforms all baseline methods and achieves state-of-the-art (SOTA).

\begin{table}[h!]
\centering
\caption{Performance Comparison of Different Methods}
\resizebox{\textwidth}{!}{
\begin{tabular}{lcc|cc|c}
\hline
Methods & \multicolumn{2}{c|}{Organs in Abdomen MRI 2D}  & \multicolumn{2}{c|}{Instruments in Endoscopy} & Cells in Microscopy \\
\cline{2-6}
& DSC ↑ & NSD ↑ & DSC ↑ & NSD ↑ & F1 ↑ \\
\hline
nnU-Net & 0.7450 ± 0.1117 & 0.8153 ± 0.1145 & 0.6264 ± 0.3024 & 0.6412 ± 0.3074  & 0.5383 ± 0.2657\\
SegResNet & 0.7317 ± 0.1379 & 0.8034 ± 0.1386 & 0.5820 ± 0.3268 & 0.5968 ± 0.3303 & 0.5411 ± 0.2633 \\
UNETR & 0.5747 ± 0.1672 & 0.6309 ± 0.1858 & 0.5017 ± 0.3201 & 0.5168 ± 0.3235 & 0.4357 ± 0.2572 \\
SwinUNETR & 0.7028 ± 0.1348 & 0.7669 ± 0.1442 & 0.5528 ± 0.3089 & 0.5683 ± 0.3123 & 0.3967 ± 0.2621 \\
U-Mamba\_Bot & 0.7588 ± 0.1051 & 0.8285 ± 0.1074 & 0.6540 ± 0.3008 & 0.6692 ± 0.3050 & 0.5389 ± 0.2817  \\
U-Mamba\_Enc & 0.7625 ± 0.1082 & 0.8327 ± 0.1087   & 0.6303 ± 0.3067 & 0.6451 ± 0.3104 & 0.5607 ± 0.2784\\
\hline
Ours\_bot & 0.7636 ± 0.1006 & 0.8322 ± 0.1034 & \textbf{0.6843 ± 0.3005} & \textbf{0.7001 ± 0.3046} & 0.5818 ± 0.2386 \\
Ours\_enc & \textbf{0.7747 ± 0.0950} & \textbf{0.8374 ± 0.0951} & \textbf{0.6843 ± 0.3024} & \textbf{0.7001 ± 0.3067} & \textbf{0.6036 ± 0.2435} \\
\hline
\end{tabular}}
\label{table:1}
\end{table}

Notably, both variations of xLSTM-UNet show superior performance across all datasets. Specifically, xLSTM-UNet\_Enc demonstrates the highest performance with a DSC of 0.7747 and an NSD of 0.8374 on the Abdomen MRI 2D dataset, outperforming the previous state-of-the-art (SOTA) model, U-Mamba by a significant margin. Additionally, xLSTM-UNet\_Bot achieves DSC and NSD scores of 0.7636 and 0.8322, respectively, surpassing the similarly structured U-Mamba\_Bot. Similarly, on the Endoscopy dataset, both xLSTM-UNet\_Bot and xLSTM-UNet\_Enc achieve the best DSC and NSD scores of 0.6843 and 0.7001, respectively. For the Microscopy dataset, xLSTM-UNet\_Enc and xLSTM-UNet\_Bot achieve F1 scores of 0.6036 and 0.5818, respectively, both surpassing the previous SOTA results, indicating their robustness in cell segmentation tasks.

\begin{figure}
\includegraphics[width=\textwidth]{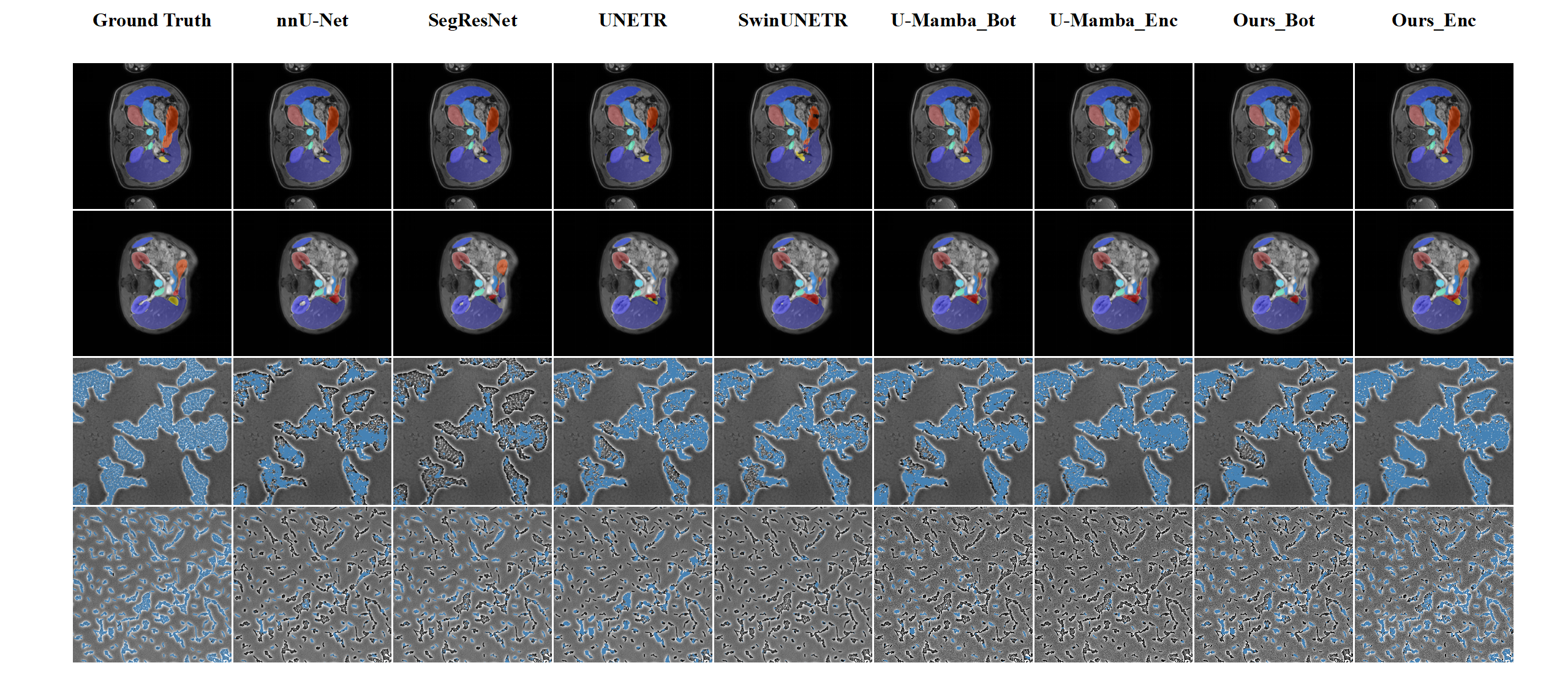}
\caption{Visualized examples of 2D medical segmentation. xLSTM-UNet demonstrates greater robustness to heterogeneous appearances and exhibits fewer segmentation errors.} \label{fig2}
\end{figure}

The visualized segmentation examples of 2D medical images further illustrate the effectiveness of xLSTM-UNet. As shown in Figure \ref{fig2}, xLSTM-UNet is more robust to heterogeneous appearances and exhibits fewer segmentation outliers compared to other models. This visual evidence underscores the quantitative results, highlighting the superior performance and reliability of xLSTM-UNet in diverse medical image segmentation tasks.

\subsection{Quantitative and Qualitative Results for 3D Segmentation}
3D medical image segmentation is generally more challenging compared to its 2D counterpart, as it involves processing a larger amount of information. The increased dimensionality leads to a dramatic surge in computational complexity, with the resolution increase causing a cubic rise in the number of computations. Accurate spatial relationship modeling is also essential for achieving satisfactory segmentation results. These factors make the xLSTM-based building blocks, with their computational efficiency, well-suited for this task.

We conducted evaluations on the 3D segmentation dataset in BraTS2023 and Abdomen MRI 3D. Table \ref{table:2} shows the performance comparison on the BraTS2023 dataset, including metrics for whole tumor (WT), tumor core (TC), and enhancing tumor (ET) regions. Our proposed method demonstrates superior performance across all evaluated metrics, including Dice and HD95, compared to other baseline methods such as SegresNet, UX-Net, MedNeXt, UNETR, SwinUNETR, SwinUNETR-V2, and SegMamba. Specifically, our method achieves the highest average Dice score of 91.80, highlighting its effectiveness in accurately segmenting brain tumor regions. Table \ref{table:3} shows the performance comparison of the Organs in the Abdomen MRI 3D dataset. Our proposed method, xLSTM-UNet\_Bot, achieves the highest Dice score of 0.8483 and the best NSD score of 0.9153, surpassing other methods such as nnU-Net, SegResNet, UNETR, SwinUNETR, and U-Mamba\_Bot. This demonstrates the robustness and accuracy of our approach in segmenting abdominal organs in MRI images.

The ability of xLSTM to effectively model semantic information in complex spatial domains has been a key factor in our success. The superior experimental results also underscore the suitability of xLSTM-based building blocks for tackling the challenge of semantic segmentation in complex imaging applications.

\begin{table}[h!]
\centering
\caption{Performance Comparison on BraTS2023 Dataset}
\begin{tabular}{lcccccccc}
\hline
Methods & \multicolumn{2}{c}{WT} & \multicolumn{2}{c}{TC} & \multicolumn{2}{c}{ET} & \multicolumn{2}{c}{Avg} \\
\cline{2-9}
& Dice↑ & HD95↓ & Dice↑ & HD95↓ & Dice↑ & HD95↓ & Dice↑ & HD95↓ \\
\hline
SegresNet & 92.02 & 4.07 & 89.10 & 4.08 & 83.66 & 3.88 & 88.26 & 4.01 \\
UX-Net & 93.13 & 4.56 & 90.03 & 5.68 & 85.91 & 4.19 & 89.69 & 4.81 \\
MedNeXt & 92.41 & 4.98 & 87.75 & 4.67 & 83.96 & 4.51 & 88.04 & 4.72 \\
UNETR & 92.19 & 6.17 & 86.39 & 5.29 & 84.48 & 5.03 & 87.68 & 5.49 \\
SwinUNETR & 92.71 & 5.22 & 87.79 & 4.42 & 84.21 & 4.48 & 88.23 & 4.70 \\
SwinUNETR-V2 & 93.35 & 5.01 & 89.65 & 4.41 & 85.17 & 4.41 & 89.39 & 4.51 \\
SegMamba & 93.61 & \textbf{3.37} & 92.65 & \textbf{3.85} & 87.71 & \textbf{3.48} & 91.32 & \textbf{3.56} \\
Ours & \textbf{93.84} & 3.89 & \textbf{92.77} & 4.06 & \textbf{88.79} & 4.14& \textbf{91.80} & 4.03\\
\hline
\end{tabular}
\label{table:2}
\end{table}

\begin{table}[h!]
\centering
\caption{Performance Comparison on Organs in Abdomen MRI 3D Dataset}
\begin{tabular}{lcc}
\hline
Methods & DSC ↑ & NSD ↑ \\
\hline
nnU-Net & 0.8309 ± 0.0769 & 0.8996 ± 0.0729 \\
SegResNet & 0.8146 ± 0.0959 & 0.8841 ± 0.0917 \\
UNETR & 0.6867 ± 0.1488 & 0.7440 ± 0.1627 \\
SwinUNETR & 0.7565 ± 0.1394 & 0.8218 ± 0.1409 \\
U-Mamba\_Bot & 0.8453 ± 0.0673 & 0.9121 ± 0.0634 \\
\hline
Ours\_bot & \textbf{0.8483 ± 0.0774 } & \textbf{0.9153± 0.0596} \\
\hline
\end{tabular}
\label{table:3}
\end{table}
\section{Discussion}
This study demonstrates that xLSTM, a model with linear computational complexity, can be an effective component in image segmentation networks. Our experimental results clearly show that xLSTM-UNet outperforms Mamba-based counterparts, underscoring the promising future of xLSTM. Given the recent huge interest in Mamba in academia, we believe that it is important to also recognize and investigate the potential of xLSTM, which has shown remarkable efficacy in this domain.

Meanwhile, medical image segmentation is inherently challenging. General image segmentation foundation models, such as Segment Anything, often fail when applied to medical images \cite{chen2023sam,chen2023sam2,kirillov2023segment}. Furthermore, medical imaging datasets are typically small. In this study, the datasets used were limited in size, which restricts our ability to explore the effects of different network scales and dataset sizes on segmentation outcomes. Investigating whether xLSTM-driven image algorithms adhere to scaling laws remains an interesting question for future research. 

Currently, xLSTM lacks dedicated optimization for hardware such as NVIDIA GPUs, which presents an opportunity for the community to contribute. Collaborative efforts are essential to optimize xLSTM for various vision tasks, leveraging its full potential. By releasing our code, we aim to encourage and facilitate further research and development, enabling the community to build on our initial findings and drive progress in this area.

This research represents an initial exploration into the application of xLSTM in medical image segmentation. There are many peaks to climb and numerous scenarios to test in this field. We hope that our comprehensive experiments and tests will demonstrate the significant potential of xLSTM in practical applications, encouraging scholars to continue exploring this promising model. With further development and optimization, we envision xLSTM achieving success comparable to that of Mamba and even Transformers, becoming a cornerstone in image segmentation and beyond.

\section{Conclusion}
In this report, we introduce xLSTM-UNet, the first U-Net architecture enhanced with Extended Long-short-memory (xLSTM) / ViL for both 2D and 3D medical image segmentation tasks. Through extensive experiments across a variety of medical imaging scenarios—including abdominal MRI, endoscopy, microscopy, and brain MRI—we have demonstrated that xLSTM-UNet significantly outperforms existing CNN-based and Transformer-based methods, as well as its Mamba-based counterparts. These findings underscore the effectiveness of xLSTM in handling complex segmentation tasks, particularly in the challenging domain of 3D medical image segmentation.

Our results show that the xLSTM-based architecture can achieve state-of-the-art (SOTA) performance, offering enhanced accuracy and efficiency. This marks a significant advancement in the field of medical image segmentation, with potential applications extending beyond healthcare. 

\begin{credits}
\subsubsection{\ackname} This research is funded by KOKONI3D, Moxin (Huzhou) Technology Co., LTD. The author thanks Qi Zhu for the discussion.

\end{credits}

%
%
%
%
\bibliographystyle{IEEEtran}
\bibliography{bib}

\end{document}